\documentclass[seceq]{jpsj2}

\begin{document}
\title{Stochastic Process Associated with Traveling Wave
Solutions\\ of the Sine-Gordon Equation}
\author{%
Tetsu \textsc{Yajima}\thanks{yajimat@is.utsunomiya-u.ac.jp}
and Hideaki \textsc{Ujino}$^{1}$\thanks{ujino@nat.gunma-ct.ac.jp}}
\inst{%
Department of Information Science, Faculty of Engineering, 
Utsunomiya University, 7-1-2 Yoto,\\ Utsunomiya, Tochigi 321-8585\\
$^{1}$Gunma National College of Technology, 580 Toriba,
Maebashi, Gunma 371-8530}
\abst{%
Stochastic processes associated with traveling wave solutions of the
sine-Gordon equation are presented.
The structure of the forward Kolmogorov equation as a conservation law
is essential in the construction of the stochastic process
as well as the traveling wave structure.
The derived stochastic processes are analyzed numerically.
An interpretation of the behaviors of the stochastic processes is given 
in terms of the equation of motion.}
\kword{stochastic process,
sine-Gordon equation,
forward Kolmogorov equation,
conservation law,
traveling wave solutions}
\maketitle
\section{Introduction}
Research on stochastic processes has received much attention 
in mathematical physics recently.
For example, the stochastic L\"owner equation\cite{refSleS}
attracts much interest regarding its connection to the conformal field
theory\cite{refSleBB}, 
as well as its original meaning as a stochastic version of
the L\"owner equation\cite{refLoewner}.
The stochastic cellular automaton and asymmetric simple exclusion
process (ASEP), among others, are also attracting considerable interest. 
The ASEP describes a non-equilibrium process of interacting particles,
whose steady state has been exactly obtained\cite{refAsepDEHP}.
Even the current at equilibrium and the corresponding phase diagram were
exactly presented for the ASEP\cite{refAsepS}.
Besides the ASEP, several stochastic cellular automata are invented
for the successful analysis of traffic flows\cite{refCAapplNFS}.
\par
The stochastic process associated with the Burgers equation is
called Burgers' process\cite{refBurgersOK},
which is applied to studies of turbulence\cite{refBurgersW}. 
This process is a typical example of nonlinear stochastic processes,
in other words, stochastic processes associated with nonlinear equations.
The time evolution of the density function of Burgers' process, for
instance, is governed  by the Burgers equation $u_u-2uu_x+u_{xx}=0$,
which is indeed nonlinear. 
Such nonlinear stochastic processes are expected to be an effective
approach to analyzing phenomena, where both nonlinear and stochastic
effects are simultaneously predominant.
\par
Among many kinds of nonlinear equations that describe various physical
systems, we have a series of integrable equations.
Expecting future applications to the analysis of phenomena related to
integrable systems, 
we shall present new nonlinear stochastic processes associated with the
sine-Gordon (SG) equation
\begin{equation}\label{eqSec01Sge}
\phi_{xt}=m^2\sin\phi,\quad
\mbox{$m$: constant},
\end{equation}
along the same line as the Burgers process.
As is well known, the SG equation is an integrable nonlinear equation
that is suitable for describing 
the dynamics of connected pendulums under gravity\cite{refSgScott},
the motions of the dislocation of one-dimensional
materials\cite{refSgDislocFK}, 
the propagation of optical waves in one-dimensional
space\cite{refSgOptical}, 
and so forth.
The SG equation possesses a kink solution that has localized structure, 
aside from periodic solutions.
In this paper, we shall consider periodic and one-kink solutions having
traveling wave structures, 
and we shall present stochastic processes associated with these
solutions. 
We shall also perform numerical analyses of the derived evolution
equations under suitable initial conditions in order to study their
behaviors.
The results and their explanation given in the later sections are
expected to describe typical features of stochastic motion associated
with integrable equations.
\par
This paper is organized as follows.
In the next section, the stochastic differential equation is presented.
A detailed description of the derived stochastic processes is given in
\S 3.
In \S 4, numerical calculations of the stochastic equations for some
kinds of traveling waves are presented, 
and interpretations of the results of the numerical analyses are shown
in \S 5.
The final section is devoted to conclusions,
including discussions on possible experimental realization.
\section{Forward Kolmogorov Equation as a Conservation Law of the 
Sine-Gordon Equation}
We shall consider a one-dimensional Ito diffusion\cite{refStochastic}, 
which is a scalar stochastic variable $X_t$ obeying the stochastic
equation
\begin{equation}\label{eqSec02ItoDiffusion}
{\rm d}X_t=b(X_t,t)\,{\rm d}t+\sigma(X_t,t)\,{\rm d}B_t,
\end{equation}
where the real-valued functions $b$ and $\sigma$ are called the drift
and diffusion coefficients, respectively.
The one-dimensional Brownian motion is denoted $B_t$.
The density function $p(x,t)$ obeys the forward Kolmogorov equation
\begin{subequations}
\begin{equation}\label{eqSec02KolForward}
\frac{\partial p}{\partial t}=A^*p,
\end{equation}
where $A^*$ is defined by
\begin{equation}\label{eqSec02ConjugateGen}
A^*=-\frac{\partial}{\partial x}\left[b(x,t)\;\cdot\;\right]
+\frac12\frac{\partial^2}{\partial x^2}\left[\sigma^2(x,t)\;\cdot\;\right].
\end{equation}
The operator $A^*$ is called the conjugate operator of the generator $A$ of
eq.\ (\ref{eqSec02ItoDiffusion}).
\end{subequations}
Equation (\ref{eqSec02KolForward}) can be interpreted as a
conservation law of $p(x,t)$, with the flux $-bp+(\sigma^2p)_x/2$.
\par
Among the conservation laws of the SG equation, 
we shall consider
\begin{equation}\label{eqSec02SgeConsLaw}
(\phi_x^2)^{}_t=(-2m^2\cos\phi)_x.
\end{equation}
Equations (\ref{eqSec02KolForward}) and (\ref{eqSec02SgeConsLaw})
have the same form when we define $p=N\phi_{x}^2$, where $N$ is the
normalization factor.
Thus, we can read
\begin{equation}\label{eqSec02Flux}
-b\phi_x^2+\frac12(\sigma^2\phi_x^2)_x^{}=-2m^2\cos\phi+A(t),
\end{equation}
where $A(t)$ is an arbitrary function.
When $\phi$ is a traveling wave solution, 
\begin{equation}
\phi=\phi(x-vt),\quad\mbox{$v$: constant,}
\end{equation}
the drift and diffusion coefficients are separated into simple forms.
Since the derivatives of the variable are given as
\begin{equation}\label{eqSec02DerivativeRelation}
\phi_x=\phi',\quad \phi_t=-v\phi',
\end{equation}
and $\phi_{xt}=-v\phi''$, we find that $-v\phi''=m^2\sin\phi$ from
eq.\ (\ref{eqSec01Sge}).
Multiplying both sides by $\phi'$ and integrating by $x$, we have
\begin{equation}\label{eqSec02Reduction}
{\phi'}^2=\phi_x^2=\frac{2m^2}{v}(\cos\phi-C),\quad
\mbox{$C$: constant}.
\end{equation}
Thus, we can eliminate $\cos\phi$ from (\ref{eqSec02Flux}) and derive
\begin{equation}\label{eqSec02FluxNew}
b\phi_x^2-\frac12(\sigma^2\phi_x^2)_x+A(t)=2m^2C+v\phi_x^2.
\end{equation}
\par
Introducing a new function $F=\sigma^2\phi_x^2$,
we can rewrite eq.\ (\ref{eqSec02FluxNew}) as
$$
b\phi_x^2=2m^2C+v\phi_x^2+\frac{F_x-A}{2}.
$$
Because $b$ and $\sigma$ should be bounded for all $x$ values even at
the zeros of $\phi_x$, we assume that $F$ and $A$ are given by
$$
F(x,t)=\frac{\alpha}{n+1}\phi_x^{n+1},\quad
A=4m^2C.
$$
Then the coefficients of eq.\ (\ref{eqSec02KolForward}) are 
determined  by
\begin{equation}\label{eqSec02CoeffsSde}
b=v+\frac{\alpha}{2}\phi_{xx}\phi_x^{n-2},\quad
\sigma^2=\frac{\alpha}{n+1}\phi_x^{n-1},\quad
n\ge2.
\end{equation}
The above restriction to $n$ prevents $b$ and $\sigma$ from diverging.
Hence, the stochastic differential equation eq.\ (\ref{eqSec02ItoDiffusion})
becomes
\begin{equation}\label{eqSec02SgSde}
{\rm d}X_t=\left(v+\frac{\alpha}{2}\phi_{xx}\phi_x^{n-2}\right)\,{\rm d}t+
\sqrt{\frac{\alpha}{n+1}}\,\phi_x^{(n-1)/2}\,{\rm d}B_t,
\end{equation}
where $v$ is the velocity of the traveling wave,
$n\ge2$, and $\alpha$ are constants.
Equation (\ref{eqSec02SgSde}) can be considered to describe the
stochastic motion of a particle moving under the influence of a
traveling wave solution of the SG equation.
When the function $\phi_x$ approaches zero exponentially,
such as the kink solution,
eq.\ (2.10) is also well-defined for $1\le n<2$,
since $\phi_{xx}\phi_x^{-1}=(\ln\phi_x)_x$.
\section{Traveling Wave Solutions of the Sine-Gordon Equation and the
Coefficient of Associated Stochastic Differential Equation}
We shall consider the coefficients in eq.\ 
(\ref{eqSec02SgSde}) for several types of solutions $\phi$ of the
SG equation (\ref{eqSec01Sge}).
In eq.\ (\ref{eqSec02Reduction}), the velocity $v$ is allowed
only negative values for $C\ge1$ and only positive values for $C\le-1$,
whereas both positive and negative $v$ values are allowed for $|C|<1$.
Among the traveling wave solutions, we choose those that have
negative velocities:
\begin{subequations}\label{eqSec03SgeTravelSols}
\begin{align}
\phi(z)&=
2\arctan\left[\sqrt{\frac{C-1}{C+1}}\,\mathop{\rm tn}\left(
\frac{\sqrt{C+1}mz}{\sqrt{-2v}},\sqrt{\frac{2}{C+1}}\right)\right],
&C>1\label{eqSec3-solution-from-tn}\\
&=2\arctan\left\{\sqrt{\frac{1-C}{1+C}}
\left[\mathop{\rm cn}\left(\frac{mz}{\sqrt{-v}},\sqrt{\frac{1+C}{2}}
\right)\right]^{-1}\right\},
&|C|<1\label{eqSec3-solution-from-cn}\\
&=4\arctan\left[e^{mz/\sqrt{-v}}\right],
&C=1.\label{eqSec3-solution-kink}
\end{align}
\end{subequations}
The functions $\mathop{\rm cn}(z,k)$ and $\mathop{\rm tn}(z,k)$ are Jacobian
elliptic functions with a modulus $k$.
We note that the solution is periodic for $C>1$ and $|C|<1$,
whereas the solution forms a solitary wave for $C=1$.
In the case associated with periodic solutions,
their supports need to be compact for these cases
or a periodic boundary condition for $x$ is required,
since the normalization factor of the density function should be finite. 
The solution eq.\ (\ref{eqSec3-solution-kink}) is the kink solution,
and we can consider the stochastic equation on the whole space.
Hereafter, the parameter $n$ is chosen to be either $n=1$ or $5$.
In the former case, its stochastic effect is spatially uniform.
In the latter case, on the other hand, its stochastic effect depends
linearly on the density function of $X_t$.
\par
\begin{figure}
\begin{center}
\includegraphics[width=85mm]{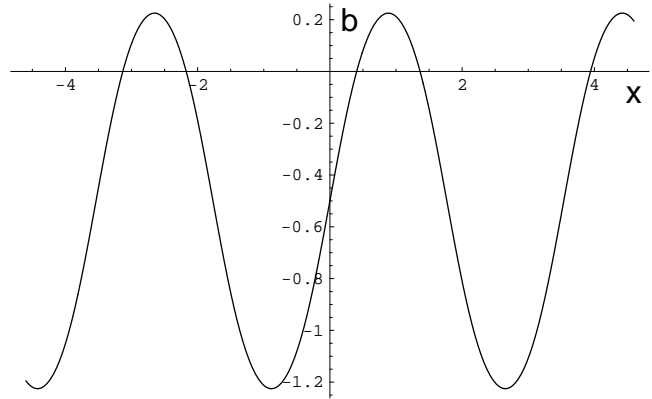}
\end{center}
\caption{Drift coefficients $b$ for the stochastic differential equation
eq.\ (\ref{eqSec02SgSde}) with $n=1$ associated with the periodic solution
(\ref{eqSec3-solution-from-tn}) for $C>1$.}
\label{figSec3First}
\end{figure}
\begin{figure}
\begin{center}
\includegraphics[width=85mm]{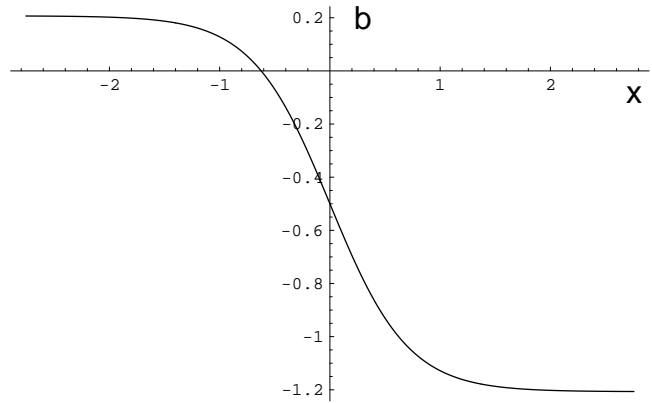}
\end{center}
\caption{Drift coefficients $b$ for the stochastic differential equation
eq.\ (\ref{eqSec02SgSde}) with $n=1$ associated with the kink solution
(\ref{eqSec3-solution-kink}), derived by setting $C=1$.}
\label{figSec3Second}
\end{figure}

In Figs.\ \ref{figSec3First} and \ref{figSec3Second}, we show two examples of
the drift coefficients $\sigma$ in the case of $n=1$,
and we display no diffusion coefficients because they are constants.
Figure \ref{figSec3First} shows a typical configuration of $b$ at $t=0$
for the periodic solution eq.\ (\ref{eqSec3-solution-from-tn}) with $C>1$.
The parameter $C$ cannot have a value at $|C|<1$, 
because periodical zeros of $\phi_x$ in the denominator of $b$
yield singularities in $b$, and the stochastic process is not well-defined.
Since $b$ is defined by eq.\ (\ref{eqSec02CoeffsSde}), 
it is periodic owing to the periodicity of $\phi$.
In the case associated with the kink solution, $C=1$, the figure of $b$
is given in Fig.\ \ref{figSec3Second}.
The coefficient $b$ has a localized structure around the center of the kink.
\par
\begin{figure}
\begin{center}
\includegraphics[width=85mm]{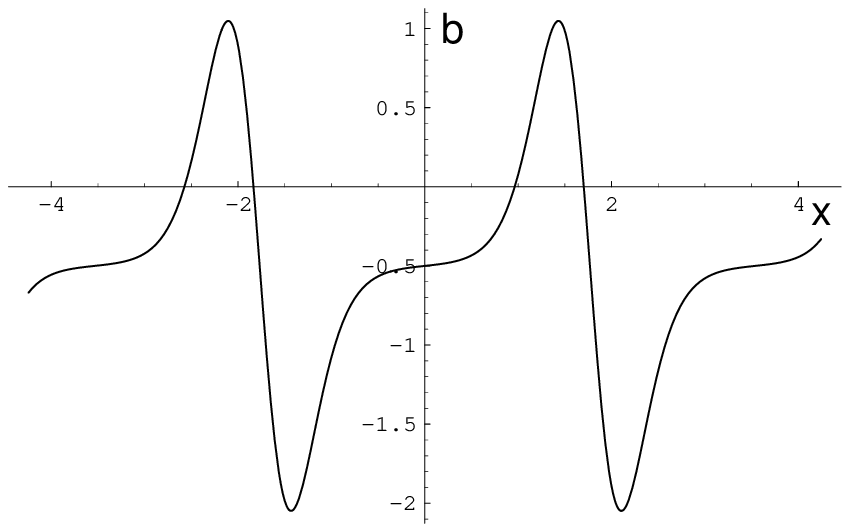}\\
(a)
\end{center}
\bigskip
\begin{center}
\includegraphics[width=85mm]{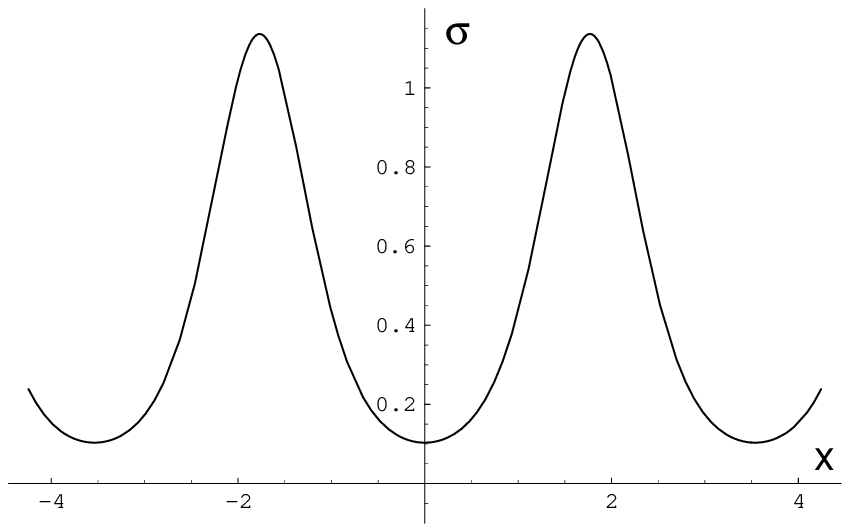}\\
(b)
\end{center}
\caption{%
Examples of the coefficients of the stochastic differential equation
(\ref{eqSec02SgSde}) with $n=5$, associated with the solution under
$C>1$. (a) The drift coefficient $b$. (b) The diffusion 
coefficient $\sigma$.}
\label{figSec3Third}
\end{figure}
\begin{figure}
\begin{center}
\includegraphics[width=85mm]{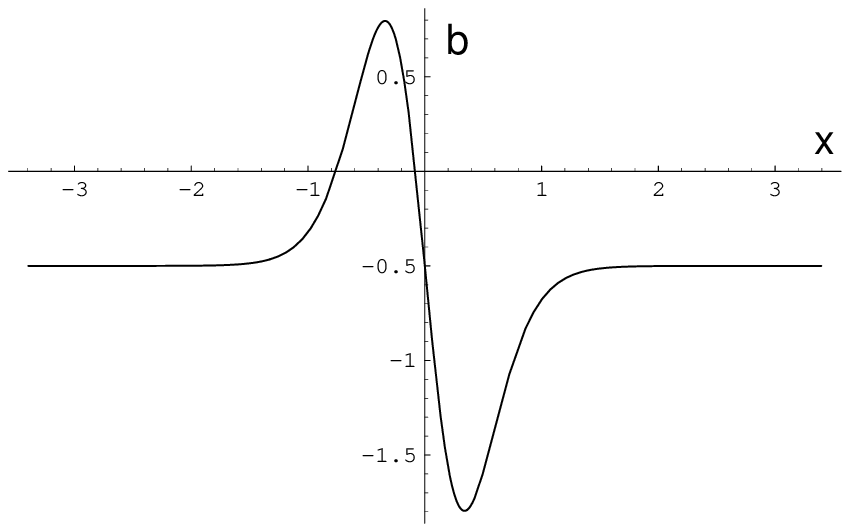}\\
(a)
\end{center}
\bigskip
\begin{center}
\includegraphics[width=85mm]{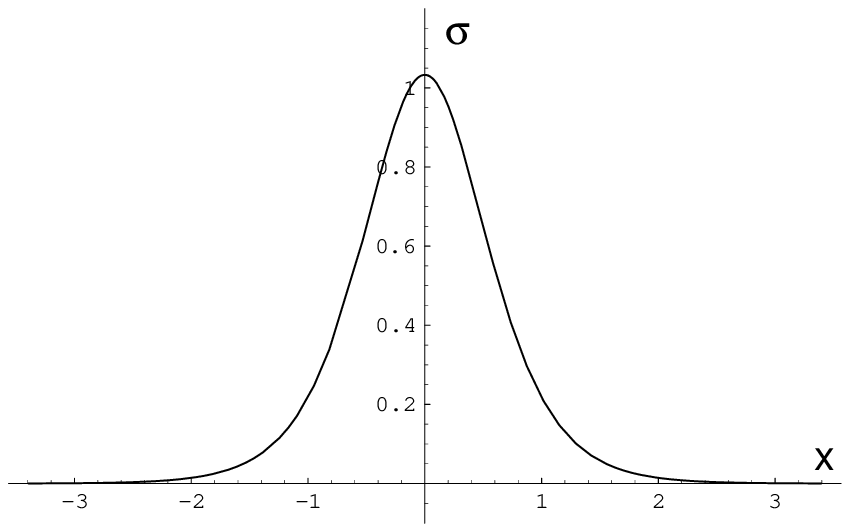}\\
(b)
\end{center}
\caption{%
Examples of the coefficients of the stochastic differential equation
eq.\ (\ref{eqSec02SgSde}) with $n=5$, associated with the kink solution
($C=1$). (a) Drift coefficient $b$. (b) Diffusion coefficient
$\sigma$.}
\label{figSec3Forth}
\end{figure}

In Figs.\ \ref{figSec3Third} and \ref{figSec3Forth}, we shall show two
typical examples of the drift and diffusion coefficients in the case of
$n=5$. 
Figure \ref{figSec3Third} shows the
coefficients for periodic solutions $C>1$.
Both $b$ and $\sigma$ have the same period, which is half of the
period of the solution $\phi$.
On the other hand, in the case of $|C|<1$,  the appearances of $b$ and
$\sigma$ are similar to those shown in the case of $C>1$, but their
period is the same as that of $\phi$.
\par
In Fig.\ \ref{figSec3Forth}, we show the coefficients of
eq.\ (\ref{eqSec02SgSde}) for the kink solution
(\ref{eqSec3-solution-kink}).
Both of the coefficients have localized structures around the center of
the kink.
Stochastic behavior of the variable $X_t$ appears only in the vicinity
of the kink, and the drift coefficient changes in the same neighborhood.
The periodicities or localized structures affect the behavior of the
sample paths of $X_t$.
\section{Numerical Results of the Stochastic Equations}
In this section, we shall analyze eq.\ (\ref{eqSec02SgSde})
numerically and study the time evolution of $X_t$, associated with the
solution eq.\ (\ref{eqSec03SgeTravelSols}).
For all the simulations, we shall take $m=1$.
The order $n$, which determines the figure of the diffusion coefficient,
is selected to be 1 and 5.
We note that all the sample paths for each case start from a fixed point,
although the initial points should be distributed in accordance with
the initial distribution $p(x,t=0)$. 
However, as we will see shortly,
the effect of this omission remains only in the early stage and
disappears after the stochastic variable $X_t$ is captured by the region
where the stochastic effect is sufficiently strong.
\subsection{Time evolution of the stochastic variable associated with
the periodic solutions for $n=1$}
In Fig.\ \ref{figSec4First}, five sample paths for the solution
eq.\ (\ref{eqSec3-solution-from-tn}) are shown together.
The parameters in $b$, $\sigma$ and $\phi$ are selected as
\begin{equation}
C=1.2,\ \alpha=1,\ v=0.5,
\end{equation}
and the initial condition is set as $X_0=0$.
The paths fluctuate around their initial points in the early stage.
Although they split into some groups of paths, the variable
$X_t$ changes along several lines parallel to each other.
The average velocity of the motion of $X_t$ is observed to be
approximately the same as that of the periodic solution.
\subsection{Time evolution of the stochastic variable associated with
the kink solution for $n=1$}
\begin{figure}
\centerline{\includegraphics[width=85mm]{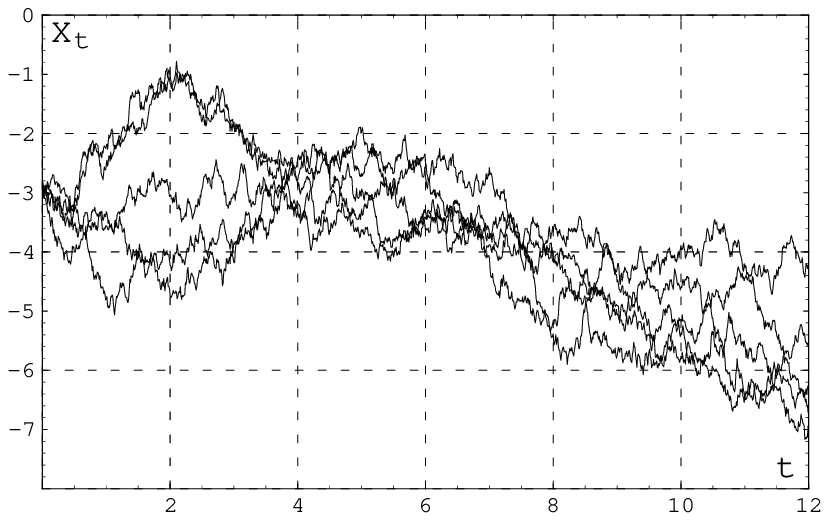}}
\caption{Sample paths for $X_t$ associated with the kink solution of
the SG equation with $n=1$.}
\label{figSec4Second}
\end{figure}

Another example is a stochastic process associated
with the 1-kink solution eq.\ (\ref{eqSec3-solution-kink}).
The coefficients are given by
\begin{equation}
b(x,t)=v-\dfrac{\alpha m}{2\sqrt{|v|}}
\tanh\dfrac{m(x-vt)}{\sqrt{|v|}}.
\end{equation}
We start with a set of parameters,
\begin{equation}\label{eqSec4-numericalparams2}
C=1,\ \alpha=1,\ v=-0.5,
\end{equation}
and the initial condition $X_0=-3$.
The results of the five paths are shown in Fig.\ \ref{figSec4Second}.
As we can see from the figure, all the paths show similar features.
The paths start from $x=X_0$, linger around the initial point 
owing to the effect of Brownian motion, and finally travel along a line
on the $X$-$t$ plane.
Although some of the paths display large deviations, 
or stay around simple values occasionally,
the average incline of each path coincides with the velocity of the peaks of the 
periodic solution of $\phi$.
\subsection{Sample paths in the case associated with the kink solution
for $n=5$}
\begin{figure}
\centerline{\includegraphics[width=85mm]{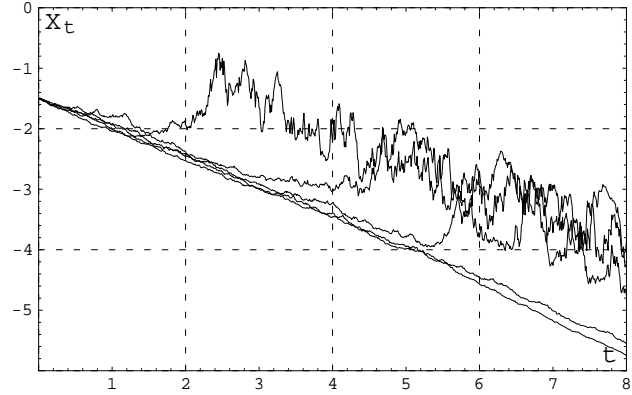}}
\caption{Sample paths for $X_t$ associated with the kink solution of
the SG equation with $n=5$.}
\label{figSec4Third}
\end{figure}

At the end of this section, we shall show the behavior of $X_t$ for
$n=5$.
Because the explicit forms of the drift coefficient for periodic
solutions in Fig.\ \ref{figSec3Third}(a) are similar to the iteration of
$b$ associated with the kink solution in Fig.\ \ref{figSec3Forth}(a),
we chose the same parameters listed in 
eq.\ (\ref{eqSec4-numericalparams2}), with the initial condition
$X_0=-1.5$.
The behaviors of $X_t$ are shown in Fig.\ \ref{figSec4Third}.
The paths initially draw curves near a simple line with small
fluctuations, and some of the paths undergo nearly the same amount of
position shifts and begin oscillating with rather large amplitudes.
In both of the initial and final states, the average incline of the
paths are observed to be the same.
Although each path of $X_t$ fluctuates intensively, and shifts from
the initial line, 
its mean path is parallel to a set of lines in the initial state.
The mean paths in the final state trace the line drawn by the center of
the kink that passes through the origin,
while the paths in the initial state have a common intercept
corresponding to $X_0$.
The time when a path falls into a state with a strong oscillation differs
from time to time.
\begin{figure}
\centerline{\includegraphics[width=85mm]{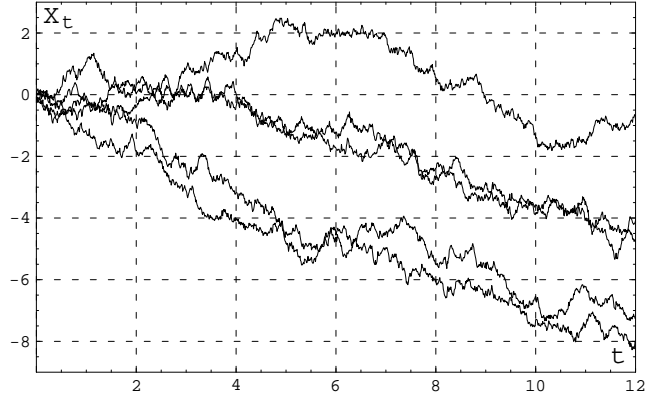}}
\caption{Sample paths for $X_t$ associated with the periodic solution of
the SG equation with $n=1$ and $C>1$.}
\label{figSec4First}
\end{figure}
\section{Asymptotic Behaviors of Sample Paths}
In the numerical results shown in the previous section, the behaviors of
the sample paths for $X_t$ in sufficiently large time regions imply the
existence of some deterministic process that drives the mean movements of
the stochastic variable, and we 
shall roughly describe this effect in terms of the equation of motion
generated from eq.\ (\ref{eqSec02SgSde}).
When we omit the second term in eq.\ (\ref{eqSec02ItoDiffusion}), we have a
nonstochastic time evolution law of $X_t$,
and we can find that the relation ${\rm d}X_t=b(X_t,t)\,{\rm d}t$ holds.
This means that the variable $X_t$ represents the position of a
particle whose velocity is $b(X_t,t)$.
Since we consider traveling wave solutions of eq.\ (\ref{eqSec01Sge})
and the drift coefficient given by eq.\ (\ref{eqSec02CoeffsSde}),
the dependences of $b(x,t)$ on space and time variables have the same
structure as those of $\phi$, which means $b=b(x-vt)$.
Differentiating the relation $\dot x=b(x-vt)$, we have the relation
\begin{equation}\label{eqSec05EomDeterministic}
\ddot x=\dfrac{{\rm d}}{{\rm d}t}b(x,t)=b'\cdot(\dot x-v)=b'\dot x-vb'.
\end{equation}
When we consider a small region around a given point of $x$, 
and approximate the graph of $b$ by its tangent whose incline is
$\beta$, eq.\ (\ref{eqSec05EomDeterministic}) reduces to
\begin{equation}\label{eqSec05EomDeterministicApprox}
\ddot x=\beta\dot x-\beta v.
\end{equation}
This equation is the same as that for a particle with resistivity
proportional to its velocity under a constant force.
For negative $\beta$, the motion results in a state of
constant velocity $v$, whereas the value of $x$ diverges for positive
$\beta$.
Thus, there appear attractors for $x$ at points
where both $b(x-vt)=v$ and $b'(x-vt)<0$ are satisfied.
\par
As an example of this explanation, let us first roughly describe the
case associated with the kink solution for $n=1$.
The exact figure of the drift coefficient is given in Fig.\ \ref{figSec3Second}.
The coefficient $b$ is almost constant in regions far from the
center of the kink, 
whereas it conspicuously decreases with $x$ around the center.
In far regions from the center of the kink, the variable $x$ moves
almost at a constant velocity toward the attractive region.
After $x$ is caught in the area where the time evolution of the variable
$x$ is approximated by eq.\ (\ref{eqSec05EomDeterministicApprox}), 
the path approaches the trajectory of the center of the kink, 
since the attractive region also travels at a velocity $v$.
\par
Next, we shall explain the behaviors of the paths for each case of the
simulation in detail. 
In the case of $n=1$ and $C>1$, there appear periodic intervals,
whose centers are the attractor of $x$.
The initial point, $X_0=0$, is one of the repulsing points, and $X_t$
travels away from the initial point at an average velocity of $b(X_t,t)$.
The motion of $X_t$ shows a small fluctuation due to the term $\sigma\,{\rm d}B_t$.
Finally, each path $X_t$ is trapped in one of the attractors and moves at
an average velocity $v$.
The three paths in Fig.\ \ref{figSec4First} are located around the
trajectories of the attractors.
Since the initial velocity of $X_t$ has a negative value, $b(X_0,0)=v<0$,
the number of paths in the region of large $X_t$ tends to be smaller
than that for smaller $X_t$.
\par
The other case of $n=1$, the result associated with the kink solution, 
can be explained in the same way.
Only a single attractor appears in this case, and all the paths fall
into the line $x=vt$ in Fig.\ \ref{figSec4Second}.
At the initial position $X_0=-3$, $b$ is approximately
$b\simeq0.2$, and $X_t$ increases gradually with stochastic
fluctuations.
After the point $t\simeq4$, the paths are caught by the attractor
near the center of the kink, and $X_t$ is driven by the attractor.
\par
In the case $n=5$ and $C=1$, we should consider the property of the
diffusion coefficient as well as the drift coefficient, because $\sigma$
is not a constant.
As in the case of $n=1$, the point $b(x-vt)=0$ with $b'<0$ plays the
role as an attractor, but the intensity of the stochastic effect is valid
only around the center of the kink.
When the $X_0$ is sufficiently far from the center, $X_t$ shows
no stochastic effect and travels at a velocity $v$,
and $X_t$ varies with little stochastic oscillation.
In the case shown in Fig.\ \ref{figSec4Third}, 
$b$ is approximately $b\simeq v=-0.5$ and $\sigma$ is rather small but
cannot be negligible near $X_0=-1.5$.
Thus, $X_t$ initially moves at a velocity $v$ approximately,
with little fluctuation.
After $X_t$ is caught by the attractor, it shifts its trajectory near
the center of the kink and begins to oscillate at a large amplitude
because of large values of $\sigma$.
The point when $X_t$ begins to oscillate is determined purely
in a stochastic way, and this causes some
of the paths to fluctuate extremely (Fig.\ \ref{figSec4Third}), but
others travel in a rather deterministic way.
\section{Discussions and Conclusions}
In this paper, we have presented new stochastic equations, associated
with the traveling wave solution of the SG equation.
By virtue of the structure of $\phi$, the unknown function of the SG equation, 
as a traveling wave,
and the assumption that the density function
$p(x,t)$ is proportional to $\phi_x$, 
we have found that the coefficients of the stochastic equation
eq.\ (\ref{eqSec02ItoDiffusion}) can be expressed in terms of the
derivatives of $\phi$, because the coefficients in the equation for
$X_t$ should not have singularities.
We have also performed numerical analyses of the time evolution of $X_t$
under some initial conditions.
The paths drawn by $X_t$ show characteristic appearances,
namely, the average $\dot X_t$ asymptotically coincides with the
velocity of the traveling wave.
These behaviors of the paths can be explained by considering attractive
points generated from the drift coefficient, which moves at the same
velocity as the solution itself.
Thus, the paths $X_t$ are observed to be driven by the solution of the SG
equation.
\par
The results derived in this paper are distinctive in that we can
consider stochastic motions of a particle, where the density function is
expressed in terms of the traveling wave solution of a solvable equation.
Since the traveling wave solutions of the SG equation are well known,
this feature enables us to derive various physical quantities related to
stochastic evolutions accurately.
In addition, we can expect several applications of the result presented
in this paper.
The behaviors of solitons under external effects such as 
impurities\cite{refIW1,refIW2}
or fluctuations in metamaterials\cite{refMetamaterial}
have attracted much interest and have been studied in various systems.
Although many results have been derived from these studies, these
studies focus on effects that act on solitons;
the stochastic effect caused by solitons has not been studied
so far.
Applying the results in this paper to the analysis of phenomena in real
materials, we can find how solitons affect original systems as
feedbacks, and how particles excited in media travel while interacting
with solitons.
\par
Since the SG equation is related to many physical systems,
the derived model can provide good descriptions of such systems 
including stochastic effects.
For example, the mechanical model of the motion of connected rotators,
which are applied forces due to twisted connection 
proportional to twist angle under gravity\cite{refSgScott},
or a dislocation in sinusoidal periodic potentials\cite{refSgDislocFK}
with heat fluctuation, 
and the optical waves in two-level atoms\cite{refSgOptical}
with heat excitations, are such candidates.
Therefore, we can expect that the stochastic processes derived in this
paper can be observed experimentally.
Such experiments in actual media or derivations of stochastic
equations from physical models are challenging problems.
\par
At the end of the paper, we would like to discuss extensions of the
derived results.
We used the special structure of traveling waves to derive the
stochastic differential equation,
namely, their coefficients $b$ and $\sigma$ in explicit forms.
Because of this restriction, the result is not applicable to the case
associated with a solution that has a time-dependent profile such as 
a two-soliton solution.
When we construct stochastic processes for general solutions,
we can consider stochastic phenomena associated with the SG equation under
various conditions.
Another extension is to consider stochastic equations for other
integrable equations.
For example, the nonlinear Schr\"odinger (NLS) equation is one of the
well-known models in integrable systems.
The method of stochastic quantization is known to be effective way for
connecting the Schr\"odinger equation with a stochastic differential equation, 
and the stochastic equation associated with the NLS equation can be used
to describe the motion of particles in various physical models such as
plasma physics or nonlinear optics.
There are many applications to be considered, and these are future
problems to be examined. 
\section*{Acknowledgment}
This work is partially supported by a Grant-in-Aid for Scientific Research
(C) $13640395$ from the Japan Society for the Promotion of Science (JSPS).

\end{document}